\documentclass[prl,twocolumn,english,superscriptaddress]{revtex4-1}

\usepackage[T1]{fontenc}
\usepackage[latin9]{inputenc}
\usepackage{color}
\usepackage{babel}
\usepackage{latexsym}
\usepackage{float}
\usepackage{amsmath}
\usepackage{amsfonts}
\usepackage{graphicx}
\usepackage{times}   
\usepackage{esint}
\usepackage{subfigure}
\usepackage{verbatim}
\usepackage[unicode=true,pdfusetitle,
 bookmarks=false,colorlinks=true,citecolor=blue,urlcolor=blue,linkcolor=red]{hyperref}

\makeatletter

\@ifundefined{textcolor}{}
{%
 \definecolor{BLACK}{gray}{0}
 \definecolor{WHITE}{gray}{1}
 \definecolor{RED}{rgb}{1,0,0}
 \definecolor{GREEN}{rgb}{0,1,0}
 \definecolor{BLUE}{rgb}{0,0,1}
 \definecolor{CYAN}{cmyk}{1,0,0,0}
 \definecolor{MAGENTA}{cmyk}{0,1,0,0}
 \definecolor{YELLOW}{cmyk}{0,0,1,0}
}

\@ifundefined{date}{}{\date{}}
\AtBeginDocument{
  
}
\makeatother

\newcommand{\SAVE}[1]{}

\begin{document}
\renewcommand\abstractname{}

\title{ A deterministic alternative to the full configuration interaction quantum Monte Carlo method}
\author{Norm M. Tubman, Joonho Lee, Tyler Y. Takeshita, Martin Head-Gordon, K. Birgitta Whaley}
\affiliation{ University of California, Berkeley, Berkeley, CA 94720, USA}
\date{\today}

\begin{abstract} 
Development of exponentially scaling methods has seen great progress in tackling larger systems than previously thought possible.  One such technique, full configuration interaction quantum Monte Carlo, is a useful algorithm that allows exact diagonalization through stochastically sampling determinants.  The method derives its utility from the information in the matrix elements of the Hamiltonian, along with a stochastic projected wave function, to find the important parts of Hilbert space.  However, the stochastic representation of the wave function is not required to search Hilbert space efficiently, and here we describe a highly efficient deterministic method to achieve chemical accuracy for a wide range of systems, including the difficult  Cr$_{2}$ dimer.  
We demonstrate that such calculations for systems like Cr$_{2}$ can be performed in just a few cpu hours.  
In addition our method also allows efficient calculation of excited state energies, for which we illustrate with benchmark results for the excited states of C$_{2}$.  
\end{abstract}
\maketitle

\newpage


{\it Introduction:}
The scope of traditional approaches to full configuration interaction (FCI) has been limited to simple diatomic molecules ~\cite{gan2005, gan2006}, and there has been little progress in diagonalizing spaces much larger than a billion determinants in recent times~\cite{olsen1996,leininger2000,olsen1996-1}.
However, recent progress in alternative approaches to FCI problems has increased the scope of FCI beyond simple diatomic molecules.  Two techniques in particular have been important in this progress, full configuration quantum Monte Carlo (FCIQMC)~\cite{booth2009}, and density matrix renormalization group (DMRG)~\cite{white1993, white1999, chan2002}.  Both algorithms provide unique advantages, with DMRG being the definitive method for systems in which one can identify degrees of freedom with low levels of entanglement~\cite{schollwock2005,stoudenmire2012}, and FCIQMC showing promise for molecules and extended systems in two or more dimensions~\cite{booth2011,shepherd2012}. The success of DMRG and FCIQMC in quantum chemistry is highlighted by their recent applications to unprecedented large-size determinant spaces while also achieving chemical accuracy~\cite{booth2013, daday2012, thomas2015, cleland2012, hachmann2007,mizukami2013,sharma2014,hitesh2012,holmes2015,kolodrubetz2013}.

The FCIQMC method is a useful technique with a few limitations which include biased sampling and comparatively computationally expensive simulations.  The biased sampling is a result of the initiator approximation~\cite{cleland2010}, generally employed in FCIQMC calculations which additionally limits the space in which determinants can be sampled.  The initiator approximation can cause errors in FCIQMC calculations that can be unexpectedly large~\cite{amaya2015}. The need for this approximation is related to the Monte Carlo sampling and not necessarily related to power of the technique which we suggest is finding important determinants.  In this Letter we suggest an alternative to stochastic sampling in favor of a completely deterministic version of the FCIQMC technique that efficiently samples the determinant space. We denote this deterministic FCIQMC algorithm by Adaptive Sampling CI (ASCI). Our approach should be contrasted to other techniques for finding energetically important determinant subspaces.  The majority of traditional CI methods encode relevant physical degrees of freedom based on excitation levels from a reference determinant ~\cite{sherrill1999}. These excitation-based methods can also suffer from inaccuracy as they miss important parts of determinant space. 
Within the CI framework several promising ways to circumvent this problem have been suggested that focus on selected CI approaches where one selects relevant determinants based on different criteria~\cite{evangelista2014,bender1969,langhoff1973,buenker1974,buenker1978,huron1973, evangelisti1983,maurizio1987,bagus1991,harrison1991,sconst1993,neese2003,roth2009,maynau2011,caffarel2015, wenjian2016,evangelista2014}. 
After presenting the ASCI method, we establish a connection between FCIQMC and various selected CI approaches. We then apply the ASCI method to the Cr$_{2}$ dimer, a classic hard problem for many computational electronic structure methods~\cite{amaya2015}.  Finally, we demonstrate that excited states are straightforward to calculate with ASCI.  We note that the calculation of excited states within stochastic FCIQMC is  possible, but require specialized techniques~\cite{booth2012}, or stochastic orthogonalization between walker sets, which is quite  different from the method described here.  

\textit{A path to a deterministic algorithm}:
In the initial development of FCIQMC, one of the original improvements on the method was to take part of the projection step and make it deterministic~\cite{hitesh2012}.  In this work, we go further and develop a completely deterministic algorithm.  Our approach here is to find important determinants in the same manner as FCIQMC, i.e. to sample determinants based on the absolute value of the ground state wave function amplitudes. The FCIQMC technique was originally presented as a projector method in imaginary time and we use this approach to motivate our method.   We start by expanding a wave function in the space of determinants,

\begin{equation}
\psi(\tau) = \sum_{i}C_{i}(\tau)|D_{i}\rangle,
\end{equation}
and the propagator in imaginary time,
\begin{equation}
-\frac{dC_{i}}{d\tau}=(H_{ii}-E)C_{i}+\sum_{i \ne j}H_{ij}C_{j},
\label{imgtime}
\end{equation}
which has an  asymptotic solution of a stationary state with $\frac{dC_{i}}{d\tau} = 0$.  In FCIQMC, the parameter $E$ is a free parameter that controls the population of walkers. Here we will consider $E$  to be the ground state energy or our best approximation thereof.  The power of FCIQMC is that it ignores the unimportant parts of determinant space, 
 finds important determinants, and samples them according to their amplitudes.  For a stationary state we can solve for the individual coefficients as $C_{i} = -\frac{\sum_{i \ne j}H_{ij}C_{j}}{(H_{ii}-E)}$.

The RHS of this equation captures all aspects of the FCIQMC algorithm.  The $H_{ij}$ in the numerator corresponds to the spawning step~(transition moves between determinants), the sign of $H_{ij}$ in the numerator and specifically summing over positive and negative terms corresponds to the annihilation step~(cancelation of positive and negative walkers), and the denominator corresponds to the death/cloning step~(adding or removing walkers from the simulation).  
The key to turning FCIQMC into a non-stochastic algorithm is to remove the stochastic sampling and replace it with a deterministic ranking of important determinants.  In both approaches the Hamiltonian matrix elements and a wave function are needed.  In FCIQMC the wave function is represented by the distribution of walkers at any given step, whereas for our deterministic algorithm, we use an approximate wave function at each iteration as follows,

\begin{equation}
C^{1}_{i} = \frac{\sum_{j \ne i}H_{ij}C^{0}_{j}}{(H_{ii}-E)}.
\label{trialwfs}
\end{equation}
In this equation, we have labeled the coefficients of the initial wave function as $C^{0}$, and the output coefficients as $C^{1}$.  Thus for any good approximation to the ground state wave function we can use Eq. \ref{trialwfs} to determine the importance of determinants in a much larger space that what is initially included in $C^{0}$. This $C^{1}$ is essentially a first-order perturbation estimate for CI coefficients in the Epstein-Nesbet perturbation theory ~\cite{huron1973}.

\textit{The Deterministic Algorithm}:  
The technique described in this section can be considered a variant of CIPSI (Configuration Interaction by Perturbation with multiconfigurational zeroth-order wave functions Selected by Iterative process)~\cite{huron1973}, with a modified search procedure.  
For the largest systems considered here, only a modest amount of memory is needed, and all parts except the diagonalization step can be trivially parallelized.    The algorithm is defined by two determinant subspaces: a core space of size \textit{cdets} and a target space of size \textit{tdets}.  The \textit{core space} determines the number of terms $j$ we include in the sum in Eq. \ref{trialwfs}.  This is to say we select $j$ determinants with the largest C$_{j}^{0}$, and consider only those to be non-zero in the sum of equation Eq. \ref{trialwfs}.  The  space to be searched is the  set of single and double excitations of the core set of determinants.  
Since our objective is to find the determinants with the largest amplitudes, we in general only need to search determinants connected to those with large amplitudes. 
We illustrate the use of this approximation with numerical tests in the next section. 
The \textit{target space} contains the top \textit{tdet} determinants, as determined  from the ranking, 
 and is the rank of the matrix diagonalized in each iteration.  

 Initalize: set size of core (\textit{cdets}) and target space (\textit{tdets}).  Set the starting wave function to the Hartree Fock wave function (C$_{0}$ = 1, C$_{i>0}=0$; E=E$_{HF}$).

(1) Evaluate the perturbed wave function amplitudes over al the single and double substitutions from the core space.   
\begin{equation}
 A_{i} =\frac{\sum_{j \ne i}^{core}H_{ij}C_{j}}{H_{ii}-E}.
\end{equation}

(2) For the core determinants of the current wave function \{C$_{i}$\}$_{core}$, and the current perturbed wave function \{A$_{i}$\}$_{search}$, select the \textit{tdet} largest absolute values to define the new target space.

(3) Form and diagonalize $H$ in the target space.

(4) The lowest eigenvalue is the new energy $E$.  The largest \textit{cdet} amplitudes, by magnitude, define the new core space.  If the energy is not converged, return to step (1).

The determinants found at the end of the simulation will in general be the most important determinants for the ground state wave function.  
 Unlike FCIQMC this technique has no population bias, initiator bias, and no sign problem.  The technique described here provides an inherently variational energy. However, it is possible to extend the accuracy of the technique with perturbation theory, which comes at the expense of the energy no longer being variational~\cite{giner2013}.

\textit{Discussion}: When performing an ASCI calculation the first few steps involve the exploration of  higher order excitations.   If the starting wave function is a single determinant, then each step increases the maximum number of excitations that have been explored by 2.  
The coefficients $A$, as calculated in step (1),  can span all single and double excitations from the current wave function, and is generally very large. Truncating the coefficients that are calculated for $A$ in the main part of the self-consistent loop allows us to consider applying our algorithm to large systems.  To maintain size consistency, the value of \textit{tdets} and \textit{cdets} will have to grow with system size.  For the systems considered in this work, it was possible to converge the value of {\it{cdets}}, as the energy with respect to this parameter can be extrapolated by running with a few different values.

Apart from the diagonalization step, the most computationally expensive task is forming the $A$ matrix and finding its largest values. This can be split among many processors, with the only non-trivial communication occurring during the final aggregation of the final values.

Further improvements are possible by generating the natural orbitals after an initial run and using them to recalculate the electron integrals. The natural orbitals can be generated at any point in the simulation from  the current best wave function.  The natural orbitals are generally thought to produce highly compact representations of a wave function~\cite{szabo:book}.
For a difficult system like Cr$_{2}$,  we find the use of natural orbitals to be  crucial to obtain accurate energies.

Before we present our results we consider numerical tests for our approximation of the matrix $A$.  In general, the coefficients of the ground state wave function will span many orders of magnitude, and searching over the determinants with the smallest amplitude coefficients is inefficient.  Using \textit{cdets} as a parameter to limit the search to only the important determinants is a well controlled approximation that can be converged. This approach has similarities to the FCIQMC initiator approximation~\cite{cleland2010}.

 We demonstrate the accuracy of this approximation by considering CN in an STO-3G basis (6240 determinants).  We show that we can find the most important determinants in Hilbert space without having to perform a diagonalization over a larger determinant space. Table \ref{tab3} provides a comparison of the most important determinants found with different values of {\it{cdets}} and {\it{tdets}} to those obtained from  a full diagonalization over the entire space.  For a simulation of 100 core determinants and a target space of 200 determinants (100/200) we found 182 of the top 200 determinants (91\%).  The remaining 9\% of missed determinants were found to be close in amplitude to the determinants that replaced them in the target space.  Similar result can be seen for for all the simulations presented. These results suggest that for some simulations that the search algorithm isn't highly dependent on the size of the core space and the small percentage of determinants that are missed by the algorithm are replaced by determinants that are similar in importance. 
Thus we argue that extremely high accuracy is not needed in determining the ranking order.  This approximation is even further reduced when perturbation corrections are used, which can correct for any important determinants that were missed.

\begin{table}[h!]
  \begin{center}
    \begin{tabular}{l|c||r||r||r||r}
      cdets & tdets & energy & top 200 & top 400 & top 800\\
      \hline
      100 & 200 & -91.17389 & 182 (91\%)  & 200  & 200  \\
      100 & 400 & -91.17637  & 200 & 373 (93\%)  & 400  \\
      100 & 800 & -91.17749  & 200  &400  & 724 (91\%)\\
      200 & 400 &-91.17657 & 200  & 387  (97\%)& 400\\
      200 & 800 &-91.17753 & 200  & 400 & 758 (95\%)\\
      400 & 800 &-91.17755 & 200  & 400 & 776 (97\%)\\
      FCI &  &-91.17767&&\\
    \end{tabular}
    \caption{Test of the CN dimer using the full search algorithm.   The energies are in units of Ha.  The columns with 'top 200; is the number of determinants we found that agree with the top 200 determinants, by amplitude, of the exact answer.   Likewise for 'top 400' and 'top 800'. The total FCI space is 6240 determinants~\label{tab3}}
  \end{center}
\end{table}
\textit{Benchmark results}:
Our main goal in presenting this analysis is to demonstrate that a deterministic method is capable of exploring determinant space in a similar manner to FCIQMC.  We did not focus on algorithmic speed and we only provide timings here to suggest an upper bound of what can be expected.  
For the results presented here we built our own implementation of this algorithm and incorporated tools from various codes~\cite{hande2014,scemama2013,kesheng2000} and electron integrals from various packages~\cite{psi4,orca}.

{\it C$_{2}$ with the cc-pVDZ basis set}:
\begin{table}[h!]
  \begin{center}
    \begin{tabular}{l|c||r||r||r||r}
      cdets & tdets & energy \\
      \hline
      4,000  & 10,000 & -75.71842 \\
      4,000  & 50,000 & -75.72626 \\
      4,000 & 100,000 & -75.72786  \\
      8,000 & 100,000 & -75.72795   \\
      4,000 & 200,000 & -75.72878 \\
      10,000 & 200,000 & -75.72891   \\
      20,000 & 200,000 & -75.72895 \\
      20,000 & 300,000 & -75.72928 \\
      15,000 & 500,000 & -75.72953 \\
      FCI &27,900,000  &-75.72985\\
    \end{tabular}
    \caption{Energy of C$_2$ molecule in units of Hartrees, at bond length 1.27273~\AA. 
 The size of the determinant space is given by the D$_{2h}$ point group and with a frozen core.  The benchmark results for this molecule with cc-pVDZ basis set is given from the following reference~\cite{leininger2000,booth2009}.~\label{c2tab} }
  \end{center}
\end{table}
For C$_{2}$ simulations, the convergence of energies to chemical accuracy was easily achieved, and neither the use of natural orbitals nor perturbation corrections are needed.  In comparison with the exact results~\cite{leininger2000}, we were able to achieve an accuracy of 1~mHa using a diagonalization no larger than 200,000 determinants.  The total computer time for a simulation of this size was less than 2 cpu hours.  Results are presented in Table \ref{c2tab}, using different values of \textit{tdets} and \textit{cdets}. 

\begin{figure}[htpb]
\centering
\includegraphics[width=1\linewidth]{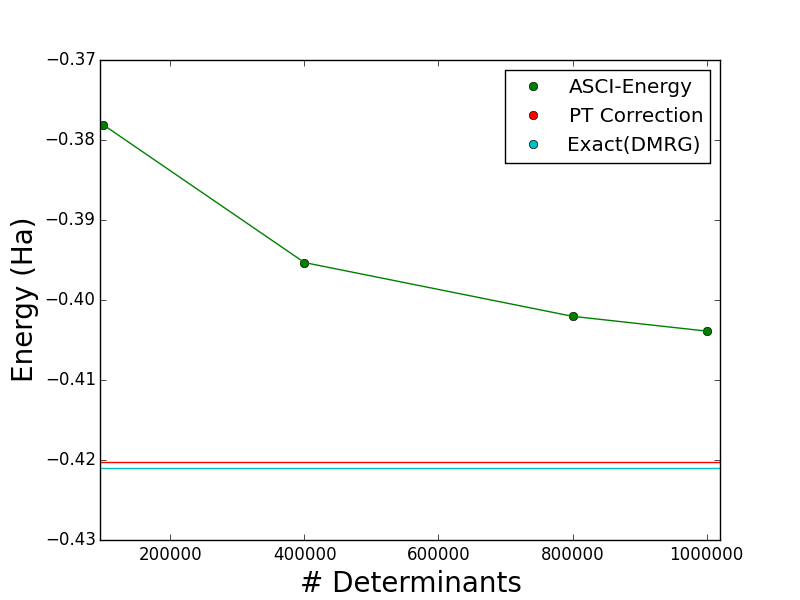}
\caption{(Color online): Cr$_{2}$ energy as a function of tdet size, with the largest target space going up to 1 million determinants. The plotted energies have been shifted by 2086 Ha. The PT correction line represents our result with the added perturbation theory correction, which brings our final energy to within 1~mHa of the predicted exact result.  Our best energy without the perturbation correction is -2086.40388~(Ha).  With the perturbation correction it is -2086.4203~(Ha).  The DMRG benchmark is -2086.420948~(Ha)~\cite{amaya2015}.}
\label{cr2ene}
\end{figure}

{\it Cr$_{2}$ with the SV basis set}:
 Fig. \ref{cr2ene} shows the convergence of our results for Cr$_{2}$.  
In order to make a comparison with previous studies~\cite{amaya2015}, Cr$_{2}$ calculations were carried out with the SV basis set~\cite{ansgar1992} at 1.5~\AA~with  24 active electrons in 30 orbitals and a frozen  core.
A compact representation of the wave function in this system is dependent on having a good set of orbitals. As part of our algorithm, we run a preliminary calculation with a target space of 100,000 determinants, after which we calculated the natural orbitals.   The resulting natural orbitals were used to recalculate the integrals for the production run.
  The total energy for our most accurate simulation converged to within 16 mHa of the predicted full CI basis set energy~\cite{amaya2015}.  

A perturbation theory analysis~\cite{giner2013} was performed bringing the final energy within 1 mHa of the predicted exact result.  
The timing for the largest simulation (\textit{tdets}=10$^{6}$), including the initial run for calculating the natural orbitals but not including the perturbation correction, requires approximately 7 cpu hours when run on a single core of a 2.40 GHz Intel Xeon processor.  While this is a relatively fast calculation, we expect further improvements will speed up the simulation significantly.  

\begin{figure}[htpb]
\centering
\includegraphics[width=0.49\linewidth]{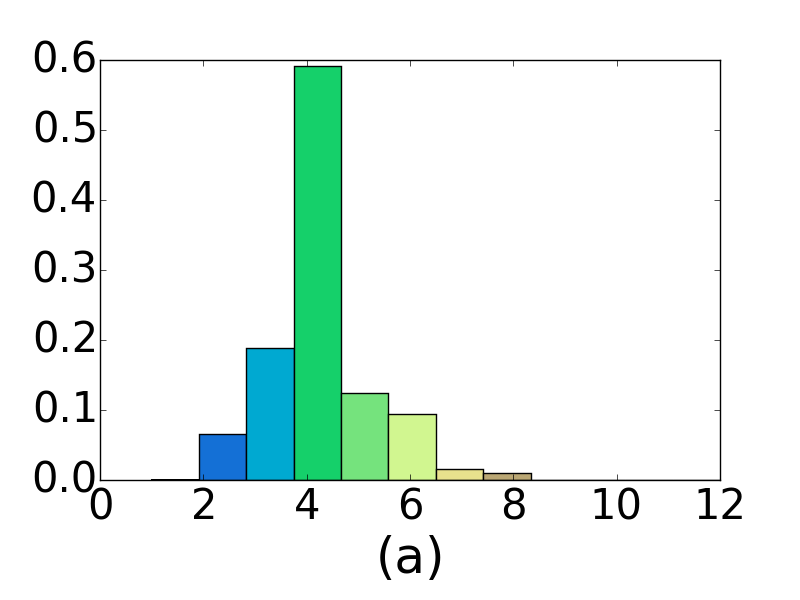}
\includegraphics[width=0.49\linewidth]{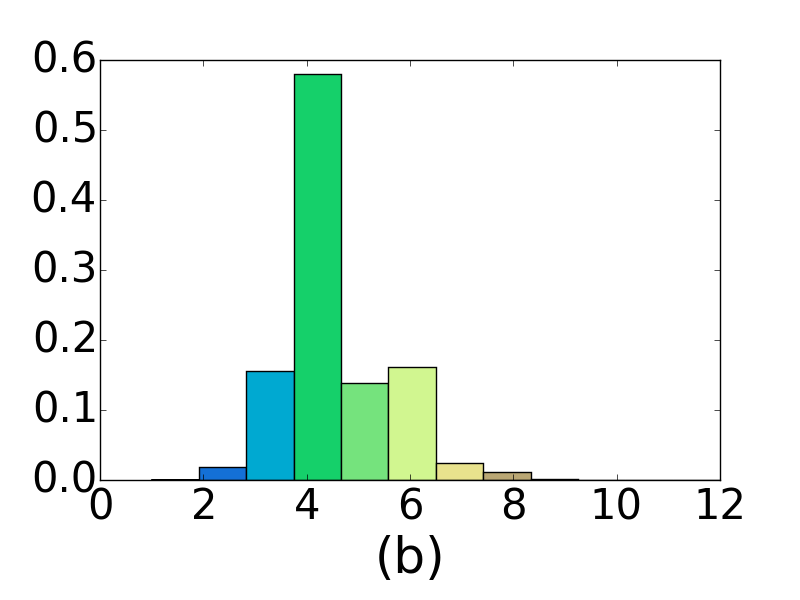}
\includegraphics[width=0.49\linewidth]{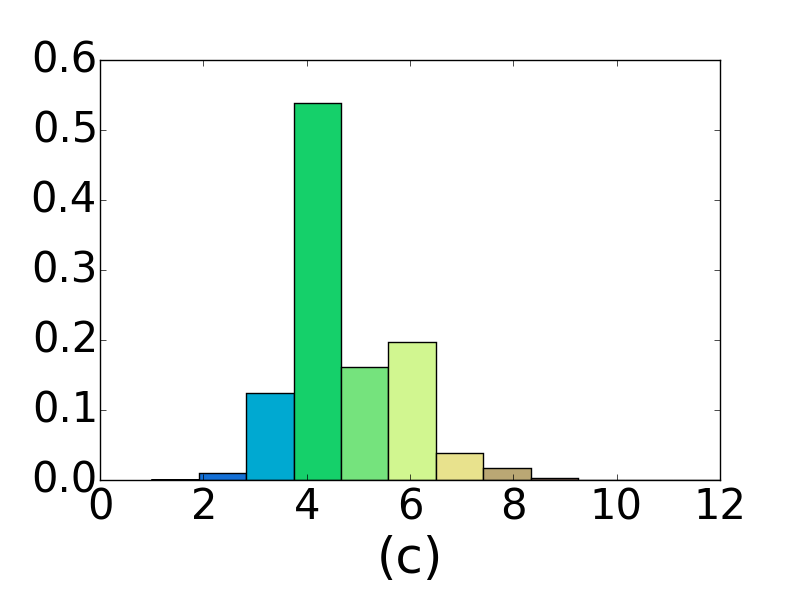}
\includegraphics[width=0.49\linewidth]{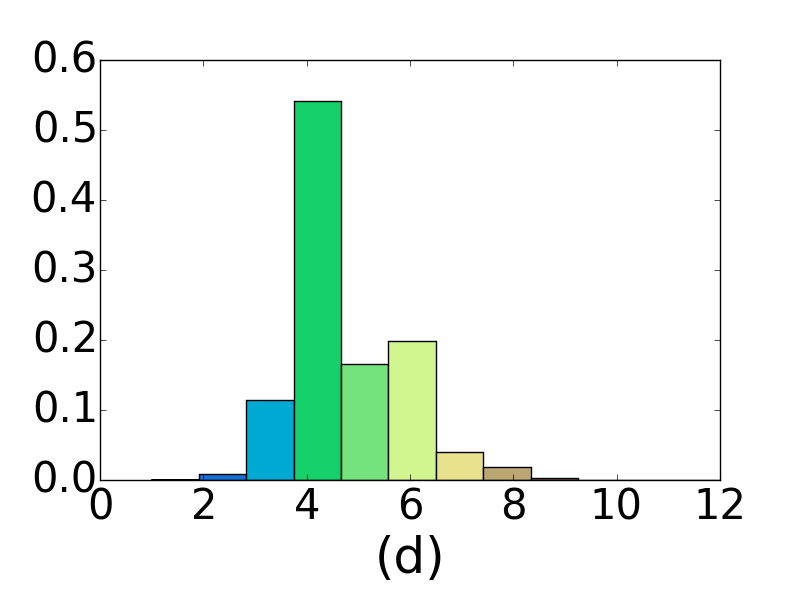}
\caption{(Color online): The distribution of determinants of Cr$_{2}$  by excitations from the dominant determinant. The x-axis is excitation level from the dominant determinant, and the y-axis is the fraction of determinants.  Plots (a), (b), (c), (d) have tdets = 100k, 400k, 800k, and 1 million respectively.   Simulations with different tdet spaces (for the ones shown here) have roughly the same fractional importance in the different excitation levels.  Thus even for our smallest tdet simulations, large determinant excitations are important. }
\label{crdist}
\end{figure}

To demonstrate the distribution of the determinants located by ASCI we plot a histogram of excitations from the dominant determinants for Cr$_{2}$ in Fig. ~\ref{crdist}.  The ratio between different excitations sectors does not change much in increasing \textit{tdet} values from 10$^{5}$ to 10$^{6}$.  For this range, the quadruple excitations generally make up half the wave function, while higher excitations above the quadruples make up roughly 30\% of the wave function.

\textit{Excited states}:
 We are also able to calculate excited states with our technique, as they are obtained automatically within the diagonalization procedure. For a 6-31G* basis of C$_{2}$, we compare against previous FCI simulations for the first two excited states~\cite{abrams2004}.  For a simulation at a distance of 1.25~\AA~with cdets=$10^{4}$ and tdets=800000, we have the following energies for the ground state and the first two excited states~($-75.7256$,$-75.6345$,$-75.6271$) in units of Ha.  The exact results are~($-75.725995$,$-75.636861$,$-75.628883$)~\cite{abrams2004}.  Thus, although the accuracy of the excited states is not as good as the ground state, it is still straightforward to achieve chemical accuracy (which is generally defined as 0.0016 Ha).  

We can improve the accuracy of excited states by noting that any eigenstate can be used in Eq. \ref{trialwfs}.  Thus we can find and rank determinants by their importance to individual excited states.  The excited state optimization can be done simultaneously with the ground state method, or in a state-by-state bootstrap method.  For the simultaneous optimization algorithm, we determine a set of important determinants to retain for each excited state.  There is a separate search step for each excited state, but one diagonalization step that combines all retained determinants.   In contrast, the bootstrap method would converge each excited state one by one, where at each step determinants would be added in specifically for the targeted excited state.  The use of natural orbitals averaged over various excited states may also be used to improve the algorithm~\cite{neese2003,shu2015}.  
A detailed study of the targeted excited state technique will be presented elsewhere.  

\textit{Connection to selected CI}:
As mentioned earlier, the ASCI method may also be considered a variant of selected CI techniques~\cite{huron1973, evangelisti1983, harrison1991, maynau2011} in which there is considerable current interest~\cite{giner2013,evangelista2014,wenjian2016,caffarel2014}. 
Our method employs first-order perturbation coefficients for selecting determinants, and is thus closely resembles the CI method, CIPSI. 
Another related technique is the $\Lambda$+SD-CI~\cite{evangelista2014} method which uses a one-step energy criteria, and a one-step approach together with our Eq. \ref{trialwfs} in order to find important determinants. Despite these similarities, none of these algorithms have been pushed to achieve chemical accuracy for hard systems, and do not appear to have been benchmarked against FCIQMC or DMRG.  The largest $\Lambda$+SD-CI calculations included roughly 50,000 determinants and attained 1--3 mHa accuracy for C$_{2}$ 6-31G* (comparable to our results in Table \ref{c2tab}).  As shown in this work, it is easy to go more than an order of magnitude in accuracy using our iterative scheme, without significantly increasing the computational effort.  The largest selected CI techniques we are aware of have been extended up to 4 million determinants~\cite{stampfub2005}, but for systems in which no benchmarks exist.

For the future development of ASCI and other selected CI techniques, it is important to consider how such methods are different from standard CI methods.  The difference is largely due to the construction of the Hamiltonian.  
Selected CI techniques need unique data structures in order to construct the Hamiltonian efficiently~\cite{stampfub2005,szalay2012}.  
A previous study demonstrated that much larger scale simulations, than what we presented here, is possible for selected CI techniques~\cite{stampfub2005}.  We are currently considering various data structures used previously~\cite{engels2001,stampfub2005} and new structures, to determine the best way to scale up our simulations.  

\textit{Conclusions}:
We have shown that the underlying dynamics of FCIQMC can be used to generate a deterministic algorithm that can be efficiently used to calculate both ground and excited states of chemical systems.  We have applied this technique to a known difficult problem in electronic structure theory, the Cr$_{2}$ molecule, and shown that chemical accuracy can be achieved with the cpu power available on any modern computer. Our results suggest that the ASCI method  (and selected CI methods in general) should be considered as a state of the art CI method in both accuracy and efficiency.  It will be interesting to determine where ASCI stands in relationship to DMRG and FCIQMC, as all these methods have different strengths.  Certainly the use of ASCI and FCIQMC is currently important since DMRG and post-DMRG methods are not yet well suited for simulations in two and three dimensions.  The ASCI method also distinguishes itself from FCIQMC in that excited states and other properties, such as the 2-RDM, are inherently easy to calculate~\cite{thomas2015-a,blunt2015}.  
  

\section{Acknowledgements}
NMT would like thank Hitesh Changlani, Cyrus Umrigar, Adam Holmes, Bryan O'Gorman, Bryan Clark,  and Jonathan Moussa for useful discussions.
This work was supported through the Scientific Discovery through
Advanced Computing (SciDAC) program funded by the U.S. Department of
Energy, Office of Science, Advanced Scientific Computing Research and
Basic Energy Sciences. We used the Extreme Science and Engineering Discovery
Environment (XSEDE), which is supported by the National Science Foundation Grant No. OCI-1053575 and
resources of the Oak Ridge Leadership Computing Facility (OLCF) at the Oak Ridge National Laboratory,
which is supported by the Office of Science of the U.S.
Department of Energy under Contract No.  DE-AC0500OR22725.

\end{document}